\documentclass[12pt]{article}
\usepackage[a4paper, total={7in, 10in}]{geometry}
\usepackage[parfill]{parskip}
\usepackage{physics, tensor, float, subcaption}
\usepackage{graphicx}
\graphicspath{ {Plots/} }
\usepackage{jhep-mod}
\usepackage{bm}
\usepackage{soul}
\usepackage{amssymb,amsmath,amsthm}
\usepackage{mathrsfs}
\usepackage[utf8]{inputenc}
\usepackage{enumerate}
\usepackage{bigints}
\usepackage{xcolor}
\usepackage{appendix}
\usepackage{graphicx}
\usepackage{float}
\usepackage{tikz}
\usepackage{setspace}
\usepackage{cancel}
\usepackage{array}
\usepackage{tabulary}
\usepackage{doi}
\definecolor{purple}{rgb}{1,0,1}
\definecolor{lime}{HTML}{A6CE39} 

\definecolor{lime}{HTML}{A6CE39}
\newcommand{\orcidicon}{%
	\begin{tikzpicture}
	\draw[lime, fill=lime] (0,0) 
		circle [radius=0.16] 
		node[white] {{\fontfamily{qag}\selectfont \tiny ID}};
	\draw[white, fill=white] (-0.0625,0.095) 
		circle [radius=0.007];
	\end{tikzpicture}
	\hspace{-5mm}
}
\newcommand\orcidJosh{{\href{https://orcid.org/0000-0003-1200-7261}{\orcidicon}}}
\newcommand\orcidMatt{{\href{https://orcid.org/0000-0003-1088-6485}{\orcidicon}}}

\renewcommand{\O}{\mathcal{O}}

\newcommand{\be}{\begin{equation}}
\newcommand{\ee}{\end{equation}}

\begin{document}


\title{\vspace{-25pt}\huge{
Explicit formulae for surface gravities in stationary circular axi-symmetric spacetimes.
}}


\author{
\Large
Joshua Baines\!\orcidJosh {\sf  and} Matt Visser\!\orcidMatt}
\affiliation{School of Mathematics and Statistics, Victoria University of Wellington, 
\\
\null\qquad PO Box 600, Wellington 6140, New Zealand.}
\emailAdd{joshua.baines@sms.vuw.ac.nz}
\emailAdd{matt.visser@sms.vuw.ac.nz}

\abstract{
\vspace{1em}

Using minimalist assumptions we develop a natural functional decomposition for the spacetime metric, and explicit tractable formulae for the surface gravities,  in arbitrary stationary circular (PT symmetric) axisymmetric spacetimes. We relate rigidity results, (the existence of a Killing horizon),  and the zeroth law to the absence of curvature singularities at the would-be horizons. 
These observations are of interest to both observational astrophysicists (modelling the cold, dark, heavy objects at the centre of most spiral galaxies),  and to the analogue spacetime community, (wherein the presence of naked singularities is not necessarily deprecated, and the occurrence of non-Killing horizons is relatively common). 

\bigskip
\noindent
{\sc Date:} 16 May 2023; 29 May 2923; \LaTeX-ed \today

ADM decomposition; null hypersurfaces; 
near horizon-physics;  on-horizon null curves; off-horizon null curves; singularities; 
rigidity conditions;  Killing horizons; \\
surface gravity; zeroth law;  horizon areas.

\bigskip
\noindent{\sc PhySH:} 
Gravitation
}

\maketitle
\def\tr{{\mathrm{tr}}}
\def\diag{{\mathrm{diag}}}
\def\cof{{\mathrm{cof}}}
\def\pdet{{\mathrm{pdet}}}
\def\QED{ {\hfill$\Box$\hspace{-25pt}  }}
\parindent0pt
\parskip7pt

\clearpage
\section{Introduction}

When modelling the cold, dark, heavy objects observed~\cite{EHT-1,EHT-4,EHT-5, EHT-SgA*-1,EHT-SgA*-6} to be present at the cores of most spiral galaxies, the Kerr black hole~\cite{Kerr, Kerr-Texas, Newman:1965, Boyer:1966, Carter:1968, Bardeen:1970, Israel:1970, Robinson:1975, Doran:1999, kerr-book-1, kerr-book-2, kerr-intro, Kerr:discovery, Teukolsky:2014, Adamo:2014, Herdeiro:2014, Herdeiro:2015, Ansatz, Darboux, unit-lapse, PG-versus-Kerr} is an excellent zeroth-order approximation. Moving beyond the zeroth order Kerr paradigm can be done either in a ``top down'' fashion, (distorting Kerr in some controlled manner~\cite{Papadopoulos:2020, Papadopoulos:2018, Benenti:1979, Carson:2020, PGLT1,PGLT2,PGLT3,PGLT4, Mazza:2021, Franzin:2022, Islam:2021,Johannsen:2015b, Johannsen:2013a, Konoplya:2016}), or in a ``bottom up'' fashion, (starting from minimal symmetry assumptions).  In recent work we have investigated a particularly attractive 3-function distortion of Kerr~\cite{3-function} (see also related considerations in references~\cite{Eye:1, Eye:2, Ghosh:2014}). Herein we shall instead work backwards from minimal symmetry assumptions:
\begin{itemize}
\item Stationarity (time independence);
\item Axi-symmetry (azimuthal independence);
\item Circularity ($\phi$-$t$ symmetry; PT symmetry).
\end{itemize}
Using these symmetries we shall first develop a canonical form for the spacetime metric that is a refinement of the usual Boyer--Lindquist form:
\begin{equation}
g_{ab}(r,\chi) = \left[\begin{array}{cc|cc}
{h_{rr}(r,\chi)\over\Delta(r)}&0&0&0\\
0 &g_{\chi\chi} (r,\chi) &0&0\\
\hline
0&0&g_{\phi\phi}(r,\chi)&-g_{\phi\phi}(r,\chi)\;\omega(r,\chi)\\
0&0&-g_{\phi\phi}(r,\chi)\;\omega(r,\chi)&\;\; - h_{tt}(r,\chi)\,\Delta(r) + g_{\phi\phi}(r,\chi)\;\omega^2(r,\chi)\\
\end{array}\right].
\end{equation}

Here  the functions $h_{rr}(r,\chi)$, $g_{\chi\chi}(r,\chi)$, $g_{\phi\phi}(r,\chi)$, and $h_{tt}(r,\chi)$ are all positive, and the presence of would-be horizons is encoded in the zeros of $\Delta(r)$. The hypersurfaces $r_H: \Delta(r)=0$ are then null hypersurfaces and so are appropriate horizon candidates.  These would-be horizons will be shown to be Killing horizons (and so exhibit a rigidity result) if and only if the Ricci scalar $R$ is finite on the would-be horizon.
Furthermore, the surface gravities of these would-be horizons will be shown to be given by the simple, tractable,  and robust formula:
\begin{equation}
\kappa = {\Delta'(r_H)\over2}
\; \sqrt{h_{tt}(r_H,\chi)\over h_{rr} (r_H,\chi)}.
\end{equation}
The zeroth law will then be shown to hold if and only if both the Ricci scalar $R$ and the quadratic curvature scalar $R^{ab} R_{ab}$ are finite on the would-be horizon. 

\enlargethispage{40pt}
In an astrophysics setting there are excellent physical reasons for forbidding naked singularities~\cite{Penrose:1973, Penrose:1999, Brady:1998, Gubser:2000, Hod:2008}, and plausible reasons for forbidding extended singularities, 
so one is naturally led to Killing horizons and the zeroth law.
In contrast in ``analogue spacetimes''  (where the ``effective metric'' governs the propagation of perturbations around some background~\cite{LRR, Acoustic, Fischer:2001, Barcelo:2001, Weinfurtner:2010, Visser:2010, Liberati:2005, Fagnocchi:2010,Schuster:2018,Fischer:2023}) there typically is no pressing need to forbid naked singularities, and non-Killing horizons are common~\cite{non-Killing}, as are formally infinite surface gravities~\cite{Liberati:2000, Fischer:2022} and violations of the zeroth law.

\clearpage
\section{From axisymmetry to Boyer--Lindquist coordinates}

To start the analysis, let us label and order the spacetime coordinates as  $(r,\chi,\phi,t)$.
 Here the coordinate $\chi=\cos\theta\in[-1,+1]$ is used in preference to the more usual choice $\theta\in[0,\pi]$ solely to improve calculational performance with symbolic algebra packages, (\emph{e.g.}, {\sf Maple}), since working with trigonometric functions tends to be a computationally expensive exercise~\cite{kerr-intro}.

Based solely on stationary axi-symmetry, (plus the mild condition that the stationary and axial Killing vectors commute), the spacetime line element  can be cast into the form~\cite{Hartle,Carroll,Wald,Weinberg,Hobson,MTW}:
\begin{equation}
ds^2 = g_{ab}(r,\chi)\; dx^a \,d x^b.
\end{equation}
The key point here is that the metric components are independent of the coordinates $t$ and $\phi$. In these coordinates the relevant Killing vectors are $T^a = (0,0,0,1)$ and $\Phi^a = (0,0,1,0)$, with $[T,\Phi]=0$.

If we now add the condition of ``circularity"~\cite{circularity,Astrid}, also commonly referred to as the ``$t$-$\phi$ orthogonality condition''~\cite{Wald:LRR}, then the spacetime metric can be cast into the block diagonal  form:
\begin{equation}
g_{ab}(r,\chi) = \left[\begin{array}{cc|cc}
g_{rr}(r,\chi)&g_{r\chi}(r,\chi)&0&0\\
g_{r\chi}(r,\chi) &g_{\chi\chi} (r,\chi) &0&0\\
\hline
0&0&g_{\phi\phi}(r,\chi)&g_{t\phi}(r,\chi)\\
0&0&g_{t\phi}(r,\chi)&g_{tt}(r,\chi)\\
\end{array}\right].
\end{equation}
In the particle physics community this symmetry would more commonly be referred to as PT symmetry,
whereby reversing the sense of rotation $\phi\to-\phi$ is equivalent to reversing the flow of time $t\to-t$.
(Equivalently, looking at the object in a mirror is the same as reversing the flow of time.)  While PT symmetry (and CPT symmetry)  are  most often discussed for QFTs in flat Minkowski space, generalizations to curved spacetime have been developed, for instance,  in reference~\cite{Hollands:2002}. 

\enlargethispage{10pt}
At this stage we are still completely free to make arbitrary 2-dimensional coordinate transformations in the $(r,\chi)$ plane. 
Let us define the $2\times2$ determinant
\begin{equation}
D(r,\chi) = \det \left[\begin{array}{cc}
g_{\phi\phi}(r,\chi)&g_{t\phi}(r,\chi)\\
g_{t\phi}(r,\chi)&g_{tt}(r,\chi)\\
\end{array}\right].
\end{equation}
which is by construction a scalar function on the  $(r,\chi)$ plane. 
We can now use the zero sets of the determinant $D(r,\chi)$ to aid in choosing suitable coordinates  on the  $(r,\chi)$ plane. 
Let us (without significant loss of generality) assume the zeros of $D(r,\chi)$ define non-intersecting lines in the $r$-$\chi$ plane, choose to label these zeros by specific values of the coordinate $r_i$, and then choose $r(x)$ in the rest of the $r$-$\chi$ plane to smoothly interpolate between the $r_i$. 

Thence, we can without loss of generality choose the coordinate $r$ to be constant on the zero sets of the $\phi$-$t$ determinant $D(r,\chi)$.
Specifically, without loss of generality we can set
\begin{equation}
D(r,\chi) = - D_0(r,\chi) \; \Delta(r),
\end{equation}
where $D_0(r,\chi)>0$ is non-zero and $\Delta(r)=0$ whenever the determinant $D(r,\chi)$ vanishes.

We can now additionally choose the remaining $\chi$ coordinate to be orthogonal to $r$. 
That is, choose the integral curves of the vector $\partial_\chi$ to be perpendicular to the level sets of $\Delta(r)$.
 This now implies
\begin{equation}
g_{ab}(r,\chi) = \left[\begin{array}{cc|cc}
g_{rr}(r,\chi)&0&0&0\\
0&g_{\chi\chi} (r,\chi) &0&0\\
\hline
0&0&g_{\phi\phi}(r,\chi)&g_{t\phi}(r,\chi)\\
0&0&g_{t\phi}(r,\chi)&g_{tt}(r,\chi)\\
\end{array}\right].
\end{equation}
So far, these are just  a minor variant of the standard Boyer--Lindquist coordinates, which we hereby see are generic to stationary axisymmetric spacetimes exhibiting PT symmetry. But considerably more can be done by more careful use of the function $\Delta(r)$ introduced above, which is why we have been so careful getting to this point. The key extra step is to globally impose Lorentzian signature.

\section{From Boyer--Lindquist coordinates to canonical metric}

To globally preserve Lorentzian signature we would want the metric determinant $\det(g_{ab}) < 0$ to be nonzero and negative throughout the spacetime. But this means that the \emph{signs} of the $2\times2$ determinants
\begin{equation}
 \det\left[\begin{array}{cc}
g_{rr}(r,\chi)&0\\
0&g_{\chi\chi}(r,\chi)\\
\end{array}\right] = g_{rr}(r,\chi) \; g_{\chi\chi}(r,\chi)
\end{equation}
and
\begin{equation}
\det \left[\begin{array}{cc}
g_{\phi\phi}(r,\chi)&g_{t\phi}(r,\chi)\\
g_{t\phi}(r,\chi)&g_{tt}(r,\chi)\\
\end{array}\right] = - D_0(r,\chi) \Delta(r)
\end{equation}
must be anti-correlated --- in particular any zero in the $2\times2$  $\phi$-$t$ determinant $D(r,\chi)=-D_0(r,\chi)\,\Delta(r)$ must correspond to a pole in the $2\times 2$   $r$-$\chi$ determinant.  This forces us to write
\begin{equation}
g_{rr}(r,\chi) = {h_{rr}(r,\chi)\over\Delta(r)}; \qquad \hbox{where} \qquad h_{rr}(r,\chi) > 0.
\end{equation}

Furthermore, based solely on the ADM decomposition, we must be able to write
\begin{equation}
g_{ab}(r,\chi) = \left[\begin{array}{cc|cc}
g_{rr}(r,\chi)&0&0&0\\
0 &g_{\chi\chi} (r,\chi) &0&0\\
\hline
0&0&g_{\phi\phi}(r,\chi)&g_{t\phi}(r,\chi)\\
0&0&g_{t\phi}(r,\chi)&\;\;-\Psi(r,\chi) + g_{t\phi}(r,\chi)^2/g_{\phi\phi}(r,\chi)\\
\end{array}\right].
\end{equation}
where $\Psi(r,\chi) = N(r,\chi)^2 > 0$ in the domain of outer communication. 
But this now implies
\begin{equation}
\Psi(r,\chi) =  h_{tt}(r,\chi)\,\Delta(r); \qquad \hbox{where} \qquad  h_{tt}(r,\chi)>0.
\end{equation}
Overall we now have 
\begin{equation}
g_{ab}(r,\chi) = \left[\begin{array}{cc|cc}
{h_{rr}(r,\chi)\over\Delta(r)}&0&0&0\\
0 &g_{\chi\chi} (r,\chi) &0&0\\
\hline
0&0&g_{\phi\phi}(r,\chi)&g_{t\phi}(r,\chi)\\
0&0&g_{t\phi}(r,\chi)&\;\; - h_{tt}(r,\chi)\,\Delta(r) +{[g_{t\phi}(r,\chi)^2]\over g_{\phi\phi}(r,\chi)}\\
\end{array}\right].
\end{equation}
Thence
\begin{equation}
D(r,\chi) = - g_{\phi\phi}(r,\chi) \; h_{tt}(r,\chi) \; \Delta(r),
\end{equation}
and
\begin{equation}
\det(g_{ab}) = -\;h_{rr}(r,\chi) \;g_{\chi\chi}(r,\chi) \; g_{\phi\phi}(r,\chi) \; h_{tt}(r,\chi) <0 .
\end{equation}

With some foresight, let us now write
\begin{equation}
g_{t\phi}(r,\chi)=-\omega(r,\phi) \; g_{\phi\phi}(r,\chi).
\end{equation}
Then for the spacetime metric we can finally write
\begin{equation}
g_{ab}(r,\chi) = \left[\begin{array}{cc|cc}
{h_{rr}(r,\chi)\over\Delta(r)}&0&0&0\\
0 &g_{\chi\chi} (r,\chi) &0&0\\
\hline
0&0&g_{\phi\phi}(r,\chi)&-g_{\phi\phi}(r,\chi)\;\omega(r,\chi)\\
0&0&-g_{\phi\phi}(r,\chi)\;\omega(r,\chi)&\;\; - h_{tt}(r,\chi)\,\Delta(r) + g_{\phi\phi}(r,\chi)\;\omega^2(r,\chi)\\
\end{array}\right].
\end{equation}

This now is our final canonical form for the spacetime metric --- making maximal use of the relevant symmetries to isolate as much of the (coordinate singularity)  behaviour as possible into the single function $\Delta(r)$, 
while allowing us to keep $h_{rr}(r,\chi)$, $g_{\chi\chi}(r,\chi)$, $g_{\phi\phi}(r,\chi)$, and $h_{tt}(r,\chi)$, well behaved and positive at the zeros of $\Delta(r)$. 

\clearpage
For the metric determinant we still have
\begin{equation}
D(r,\chi) = - g_{\phi\phi}(r,\chi) \; h_{tt}(r,\chi) \; \Delta(r),
\end{equation}
and
\begin{equation}
\det(g_{ab}) = -\;h_{rr}(r,\chi) \;g_{\chi\chi}(r,\chi) \; g_{\phi\phi}(r,\chi) \; h_{tt}(r,\chi) <0 .
\end{equation}
We emphasise that the canonical form deduced above is indeed sufficiently general to easily accomodate essentially all of the recently introduced  quasi-phenomenological modifications of Kerr that have been the subject of so much recent scrutiny. See for instance references~\cite{Papadopoulos:2020, Papadopoulos:2018, Benenti:1979, Carson:2020, PGLT1,PGLT2,PGLT3,PGLT4, Mazza:2021, Franzin:2022, Islam:2021,Johannsen:2015b, Johannsen:2013a, Konoplya:2016} 
and~\cite{Broderick:2013, Psaltis:2014, Cardoso:2016, Bambi:2011, Bambi:2016, Barausse:2008,  
Carballo-Rubio:2018a, Carballo-Rubio:2018b, Lima:2020a, Shaikh:2021, 
Carballo-Rubio:2022, Johannsen:2015a, Johannsen:2013, Franzin:2021, KSZ-3, Simpson:2018, Lobo:2020, STU, Medved:2004}.

\section{Locating the would-be horizons}

Our primary ansatz for locating the would-be horizons will be to make use of the result
\begin{equation}
    g_{rr}(r,\chi)={h_{rr}(r,\chi)\over\Delta(r)}.
\end{equation}
(We again emphasize that this is not  a constraint, it is ultimately a coordinate choice.)

{\bf Lemma:} Under this ansatz the roots of $\Delta(r)=0$ define null hypersurfaces that trap geodesics --- and so exhibit two key features of what we want horizons to be.

{\bf Proof} (part 1): \\
Note
\begin{equation}
g^{ab} \; \partial_a r\; \partial_b r = g^{rr} = {\Delta(r)\over h_{rr}(r,\chi)} \to 0.
\end{equation}
So the constant-$r$ surfaces defined by the roots of $\Delta(r)=0$ have a null normal $\partial_a r$; 
they are indeed null hypersurfaces as desired.
\QED

{\bf Proof} (part 2): \\
Let $i,j\in\{\chi,\phi,t\}$ then 
\begin{equation}
\Gamma^r{}_{ij} = g^{rr} \left( g_{r(i,j)} - {1\over2} g_{ij,r} \right) 
= - {1\over2} \; {\Delta(r)\over h_{rr}(r,\chi)} \;\; g_{ij,r}.
\end{equation}
But this vanishes whenever $\Delta(r)=0$.
So any geodesic starting on a would-be horizon, $\Delta(r)=0$, and tangent to the would-be  horizon, (so that $dr/d\lambda=0$ initially), will stay there. \QED

\section{The 3-metric on the would-be horizon}

The 3-metric on the would-be horizon $\Delta(r)=0$ is then singular, with signature $\{++0\}$. Taking $i,j\in\{\chi,\phi,t\}$  we have
\begin{equation}
g_{ij} = 
 \left[\begin{array}{c|cc}
 g_{\chi\chi} (r,\chi) &0&0\\
 \hline
0&g_{\phi\phi}(r,\chi)&-g_{\phi \phi}(r,\chi)\,\omega(r,\chi)\\
0&-g_{\phi \phi}(r,\chi)\,\omega(r,\chi)&g_{\phi\phi}(r,\chi)\,\omega(r,\chi)^2\\
\end{array}\right].
\end{equation}
That is, on the would-be horizon we have:
\begin{equation}
ds^2 =  g_{\chi\chi} (r,\chi)\;d\chi^2 + g_{\phi\phi}(r,\chi)\; [d\phi -\omega(r,\chi) dt]^2.
\end{equation}
So the on-would-be-horizon null curves are simply
\begin{equation}
\ell^a\equiv (0,0,\dot\phi,1)^a= \left(0, 0, \omega(r,\chi) ,1 \right)^a.
\end{equation}
This is quite restrictive.

Alternatively: We could consider the vector $\ell^a \equiv (0,\dot\chi,\dot\phi,1)^a$,  and compute
\begin{equation}
g_{ab}\; \ell^a\, \ell^b = g_{\chi\chi}\; \dot \chi^2 + g_{\phi\phi} [\dot\phi -\omega(r,\chi)]^2.
\end{equation}
Since this is a sum of squares, for on-would-be horizon null curves we again find $\dot\chi=0$ and $\dot\phi=\omega(r,\chi)$. That is 
\begin{equation}
\ell^a = (0,0,\dot\phi,1)^a = \left(0, 0, \omega(r,\chi) ,1 \right)^a.
\end{equation}

\section{Enforcing  rigidity --- Killing horizons}
Let $K^a = (0,0,\Omega,1)$, with $\Omega$ constant,  be a general stationary axisymmetric Killing vector.
Then in the current framework
\begin{equation}
g_{ab} K^a K^b = g_{tt}(r,\chi) + 2 \Omega \, g_{t\phi}(r,\chi) + \Omega^2 \, g_{\phi\phi}(r,\chi)
=
 g_{\phi\phi}(r,\chi) [ \Omega- \omega(r,\chi) ]^2.
\end{equation}
Can we make  this vanish on the would-be horizons?

\enlargethispage{10pt}
Yes, easily so, \emph{mathematically} we can do so provided we set
\begin{equation}
\omega(r,\chi) = \omega_0(r) + \Delta(r) \; \tilde\omega(r,\chi),
\end{equation}
since then on the would-be horizons we have 
\begin{equation}
\omega(r_H,\chi)= \omega_0(r_H).
\end{equation}
Finally, we set $\Omega_H = \omega_0(r_H)$ to find a Killing vector that is null on the appropriate would-be horizon.
Note that this \emph{choice} enforces a separate rigidity condition on each would-be horizon.
Furthermore in the current framework we can enforce the existence of Killing horizons \emph{independently} of whether or not we say anything about surface gravities and/or try to enforce the zeroth law.

Is there a good \emph{physics} reason for enforcing Killing horizons?  Yes, if we are dealing with the physical spacetime metric --- Killing horizons are required to keep the Ricci scalar finite on the would-be horizons. 

Suppose the would-be horizon is non-extremal, ($\Delta'(r_H)\neq 0$), then in the vicinity of the would-be horizon one has
\begin{equation}
\Delta(r) = \Delta'(r_H) [r-r_H] + \O([r-r_H]^2). 
\end{equation}
A quick calculation then yields
\begin{equation}
R = {g_{\phi\phi}(r_H,\chi)^2\; [\partial_\chi \omega(r_H,\chi)]^2\over 
2 g_{\chi\chi}(r_H,\chi)\; h_{tt}(r_H,\chi)\; \Delta'(r_H)\; [ r-r_H]} + \O(1).
\end{equation}
Only if $\partial_\chi \omega(r_H,\chi) =0$, implying a Killing horizon with $\omega(r_H,\chi)=\Omega_H$, is the Ricci scalar finite on the would-be horizon. (And the singularity is even worse for an extremal would-be horizon where $\Delta'(r_H)\to 0$.)

Finiteness of the Ricci scalar is certainly a desirable property for the physical spacetime metric of general relativity~\cite{Penrose:1973, Penrose:1999, Brady:1998, Gubser:2000, Hod:2008}, but is less crucial for ``analogue spacetimes'' where the ``effective metric'' controlling the propagation of field theoretic perturbations can easily exhibit curvature singularities~\cite{LRR,Acoustic, Fischer:2001, Barcelo:2001,Weinfurtner:2010, Visser:2010, Liberati:2005, Fagnocchi:2010, non-Killing, Fischer:2023, Liberati:2000,Schuster:2018}.  This is the principal reason why various efforts at finding non-Killing horizons focus on variants of the ``analogue spacetime'' programme.

\section{Off-would-be-horizon null curves}

With the goal of now studying surface gravities, it is useful to determine a suitable class of \emph{off-would-be-horizon} null curves, null curves that become null geodesics on the would-be horizons. Consider
\begin{equation}
\ell^a = \left(\Delta(r)\;\sqrt{h_{tt}(r,\chi)\over h_{rr} (r,\chi)},\; 0, \; \omega(r,\chi) ,1 \right)^a.
\end{equation}
Then it is easy to check, using the canonical metric
\begin{equation}
g_{ab}(r,\chi) = \left[\begin{array}{cc|cc}
{h_{rr}(r,\chi)\over{\Delta(r)}}&0&0&0\\
0 &g_{\chi\chi} (r,\chi) &0&0\\
\hline
0&0&g_{\phi\phi}(r,\chi)&-g_{\phi \phi}(r,\chi)\,\omega(r,\chi)\\
0&0&-g_{\phi \phi}(r,\chi)\,\omega(r,\chi)&
\;-h_{tt}(r,\chi) \Delta(r)+g_{\phi\phi}(r,\chi)\,\omega(r,\chi)^2\\
\end{array}\right],
\end{equation}
 that even off the would-be horizon we have
\begin{equation}
g_{ab}\;\ell^a\,\ell^b = 0.
\end{equation}
These null curves are in general not null geodesics, but they do become null geodesics on the would-be horizons $\Delta(r)=0$.

Specifically, we can easily calculate (\emph{e.g.}, {\sf Maple})
\begin{eqnarray}
\ell^b \nabla_b \ell^a &=&
\left({ \partial_r [\Delta(r) \,h_{tti}(r,\chi)] 
\over \sqrt{h_{tt} (r,\chi) \, h_{rr}(r,\chi)} }\right)\,\ell^a
\nonumber
\\
&&
+ \left({\Delta(r) h_{rr}(r,\chi) \over 2 g_{\chi\chi}(r,\chi) } \;
\partial_\chi\left[h_{ttt} (r,\chi) \over h_{rr}(r,\chi)\right] \right) (0,1,0,0)^a.
\end{eqnarray}
On the would-be horizons, $\Delta(r)=0$, this reduces to
\begin{equation}
\ell^b \nabla_b \ell^a =2 \kappa \; \ell^a; \qquad\hbox{with}\qquad 
\kappa = {\Delta'(r_H)\over2} \; \sqrt{h_{tt}(r_H,\chi)\over h_{rr} (r_H,\chi)}.
\end{equation}
This construction yields a simple and tractable formula for the surface gravity of any stationary axisymmetric circular (PT symmetric) spacetime; a formula which we now check using various other presentations. 

\section{Surface gravities for would-be horizons}

There are a number of other slightly  differing definitions of surface gravity also in common use~\cite{non-Killing}; fortunately the most important of these definitions will agree with each other in the present context. 

\subsection{Inaffinity version of surface gravity}
From the above, the inaffinity version of surface gravity (\emph{cf.}~\cite{non-Killing}) is
\begin{equation}
\kappa_\mathrm{inaffinity} = {\Delta'(r_H)\over2}
\; \sqrt{h_{tt}(r_H,\chi)\over h_{rr} (r_H,\chi)}.
\end{equation}

\subsection{Peeling-off version of surface gravity}
The peeling-off version of surface gravity  (\emph{cf.}~\cite{non-Killing}) is determined by seeing how quickly the null curves ``peel off'' from the horizons.
Specifically, let us consider the null vector field
\begin{equation}
\ell^a = \left(\Delta(r)\; \sqrt{h_{tt}(r,\chi) \over h_{rr} (r,\chi)}
, 0, \omega(r,\chi) ,1 \right)^a 
\end{equation}
and compute
\begin{equation}
{d [r-r_H]\over dt} = \left[ {d\over dr} \left(\Delta(r)\; \sqrt{h_{tt}(r,\chi) \over h_{rr} (r,\chi)}
\right)\right]_H [r-r_H] + \O([r-r_H]^2)
\end{equation}
Comparing this with the standard ``peeling'' definition of surface gravity 
\begin{equation}
{d [r-r_H]\over dt} =  2\kappa_\mathrm{peeling} [r-r_H] + \O([r-r_H]^2),
\end{equation}
and noting that $\Delta(r)\to 0$ on the would-be horizons, 
we again have 
\begin{equation}
\kappa_\mathrm{peeling} = {\Delta'(r_H)\over2} \; \sqrt{h_{tt}(r_H,\chi)\over h_{rr} (r_H,\chi)}.
\end{equation}
(This agrees with the inaffinity notion of surface gravity.)

\subsection{Generator-based version of surface gravity}

A ``generator'' based version of surface gravity (based on reference~\cite{non-Killing})) is
\begin{equation}
\kappa_\mathrm{generator}^2 = {1\over 2} \left\{ \ell_{[a,b]} \;g^{bc}\; \ell_{[c,d]} \;g^{da}\right\}_H.
\end{equation}
This is more general than Wald's Killing vector construction~\cite{Wald,Wald:LRR}, since it only depends on having a null vector field that asymptotes to a horizon-generating null vector field on each would-be horizon. 
In the current setup a brief calculation ({\sf Maple}) yields
\begin{equation}
\kappa_\mathrm{generator}^2 = 
 {\Delta'(r_H)^2\; h_{tt}(r,\chi)\over4h_{rr} (r_H,\chi) },
\end{equation}
so that
\begin{equation}
\kappa_\mathrm{generator} =  {\Delta'(r_H)\over2}
\; \sqrt{h_{tt}(r_H,\chi)\over h_{rr} (r_H,\chi)}.
\end{equation}
the sign being chosen so that the outermost horizon has positive surface gravity. 

\subsection{Summary so far}

In the framework we have developed herein
\begin{equation}
\kappa_\mathrm{generator}  =  \kappa_\mathrm{peeling} = \kappa_\mathrm{inaffinity} 
= {\Delta'(r_H)\over2} 
\; \sqrt{h_{tt}(r_H,\chi)\over h_{rr} (r_H,\chi)}.
\end{equation}
These three definitions agree with each other, without the need for (as yet) enforcing Killing horizons, and without the need for (as yet) enforcing a zeroth law. We shall soon see how, if we desire, mathematically  we can \emph{independently} enforce both Killing horizons and the zeroth law.

\section{Enforcing the zeroth law}
If we now wish to enforce the zeroth law, then from
\begin{equation}
\kappa =  {\Delta'(r_H)\over2}
\; \sqrt{h_{tt}(r_H,\chi)\over h_{rr} (r_H,\chi)},
\end{equation}
we see that  \emph{mathematically}  it is sufficient to set
\begin{equation}
{ h_{tt} (r,\chi) \over h_{rr}(r,\chi)}= J(r)+\Delta(r) \, K(r,\chi); 
\end{equation}
since then on any of the would-be horizons we have
\begin{equation}
\kappa = {\Delta'(r_H) \over 2} \; \sqrt{ J(r_H)}.
\end{equation} 
This manifestly satisfies the zeroth law.
Note that in the current framework we have been able to impose the zeroth law \emph{without} necessarily  forcing the horizons to be Killing horizons. 

Is there a good physics reason for enforcing the zeroth law? We have already seen that near a would-be horizon the Ricci scalar satisfies
\begin{equation}
R = {g_{\phi\phi}(r_H,\chi)^2\; [\partial_\chi \omega(r_H,\chi)]^2\over 
2 g_{\chi\chi}(r_H,\chi)\; h_{tt}(r_H,\chi)\; \Delta'(r_H)\; [ r-r_H]} + \O(1).
\end{equation}

So a finite Ricci scalar requires $\partial_\chi \omega(r_H,\chi)\to 0$. 
A similar calculation applied to $R^{ab}R_{ab}$, and assuming you have already made $R$ finite,  yields
\begin{equation}
R^{ab}R_{ab}   = {\Delta'(r_H)\over 8 g_{\chi\chi}(r_H,\chi) \, h_{rr}(r_H,\chi)}  
\left( \partial_\chi[h_{tt}(r_H,\chi)/h_{rr}(r_H,\phi)]\over 
[h_{tt}(r_H,\chi)/h_{rr}(r_H,\phi)] \right)^2 \left(1\over r-r_H\right)
+ \O(1).
\end{equation}

So requiring finiteness of the quadratic scalar $R^{ab}R_{ab} $ (in addition to finiteness of the Ricci scalar)  imposes the additional requirement that $\partial_\chi[h_{tt}(r_H,\chi)/h_{rr}(r_H,\phi)]\to 0$, which in view of our general result for the surface gravity implies the zeroth law.

\section{Surface gravities for Killing horizons}
For completeness, note that if we choose all the horizons to be Killing horizons we can use Wald's trick~\cite{Wald,Wald:LRR} for calculating the surface gravity:
\begin{equation}
\kappa_\mathrm{Killing} = \left.\sqrt{ - {1\over2} (K_{H})_{a;b} \; (K_{H})^{a;b}}\right|_{r_{H}}.
\end{equation}
Using Killing's equation we can write this as
\begin{equation}
\kappa_\mathrm{Killing} 
= \left.\sqrt{ + {1\over2} (K_{H})_{a;b} \; (K_{H})^{b;a}}\right|_{r_{H}}
= \left.\sqrt{ + {1\over2} (K_{H})^a{}_{;b} \; (K_{H})^{b}{}_{;a}}\right|_{r_{H}}.
\end{equation}

Brute force (\emph{e.g.}, {\sf Maple}) now again recapitulates  the central result:
\begin{equation}
\kappa = {\Delta'(r_H) \over 2}
\; \sqrt{h_{tt}(r_H,\chi)\over h_{rr} (r_H,\chi)},
\end{equation} 
We had already derived this formula without using Killing horizons --- so this is at this stage just a consistency check.

\section{Horizon areas}

In contrast to the relatively strong statements we can make regarding the surface gravities and rigidity results, very little can be said about horizon areas. Consider any constant-$r$ constant-$t$ 2-surface spanned by coordinates $x^i=(\chi,\phi)$. Then the induced 2-metric is
\begin{equation}
g_{ij} = \left[ \begin{array}{cc} g_{\chi\chi}(r,\chi) & 0 \\ 0 & g_{\phi\phi}(r,\chi) \end{array} \right].
\end{equation}
The area is 
\begin{equation}
A(r) = 2\pi \int_{-1}^{+1} \sqrt{ g_{\chi\chi}(r,\chi) \,g_{\phi\phi}(r,\chi)} \;d\chi.
\end{equation}
In general nothing more can be said, even on horizon. Only after more precisely specifying the spacetime (for example, the specific 3-function distortions of Kerr considered in reference~\cite{3-function}) can anything more stringent be said concerning horizon areas.

\section{Conclusions}
What have we learnt from this discussion? Starting from the minimalist stationary axisymmetry and circularity  (PT symmetry) assumptions,  and ordering the coordinates as $x^a=(r,\chi,\phi,t)$, we can write the metric in the Boyer--Lindquist  form
\begin{equation}
g_{ab}(r,\chi) = \left[\begin{array}{cc|cc}
g_{rr}(r,\chi)&0&0&0\\
0 &g_{\chi\chi} (r,\chi) &0&0\\
\hline
0&0&g_{\phi\phi}(r,\chi)&g_{t\phi}(r,\chi)\\
0&0&g_{t\phi}(r,\chi)&g_{tt}(r,\chi)\\
\end{array}\right].
\end{equation}
Furthermore, taking into consideration the ADM decomposition,  without any further loss of generality we may write this metric in the canonical form
\begin{equation}
g_{ab}(r,\chi) = \left[\begin{array}{cc|cc}
{h_{rr}(r,\chi)\over{\Delta(r)}}&0&0&0\\
0 &g_{\chi\chi} (r,\chi) &0&0\\
\hline
0&0&g_{\phi\phi}(r,\chi)&-g_{\phi \phi}(r,\chi)\,\omega(r,\chi)\\
0&0&-g_{\phi \phi}(r,\chi)\,\omega(r,\chi)&\;
- h_{tt}(r,\chi) \Delta(r) +g_{\phi\phi}(r,\chi)\,\omega(r,\chi)^2\\
\end{array}\right]. 
\end{equation}

We have found that the roots of $\Delta(r)=0$ are the locations of the would-be horizons in this geometry, since $\Delta(r)=0$ defines a null hypersurface, while in addition one has $\Gamma^r{}_{ij}\propto \Delta(r)$ which  vanishes on the would-be horizon. Furthermore, \emph{mathematically} these would-be horizons can be promoted to Killing horizons so long as we set
\begin{equation}
\omega(r,\chi)=\omega_0(r)+\Delta(r)\;\Tilde{\omega}(r,\chi).
\end{equation}
This implies the rigidity result, $\omega(r_H,\chi)=\Omega_H$,  and \emph{physically} is tantamount to demanding that the Ricci scalar be finite on the would-be horizon.

Using three distinct methods, we then calculated a simple, tractable, and robust expression for the surface gravity on these horizons which is given as 
\begin{equation}
\kappa={\Delta'(r_H) \over 2}
\; \sqrt{h_{tt}(r_H,\chi)\over h_{rr} (r_H,\chi)},
\end{equation}
If we wish to enforce the zeroth law then \emph{mathematically} it is sufficient to set
\begin{equation}
 {h_{tt} (r,\chi) \over h_{rr}(r,\chi)}= J(r)+\Delta(r) \; K(r,\chi); 
\end{equation}

In this case, the surface gravity becomes constant on the horizon. It is important to note that this condition can be imposed without necessarily forcing the horizons to be Killing horizons. 

If we do want to enforce Killing horizons, then provided we set $\Omega_H=\omega_0(r_H)$, the norm of Killing vector $K^a=(0,0,\Omega_H,1)$ becomes zero, so this Killing vector becomes null on the horizons. If we do so, then a fourth method for calculating the surface gravity becomes available, and agrees with the previous three methods. 

Overall, we have seen that with minimal assumptions, stationary axisymmetry plus ``circularity''(PT symmetry) one can give a surprisingly simple expression for the surface gravity on the would-be horizons.
Additionally we can quite easily enforce a zeroth law and/or the existence of Killing horizons.
(This being related to the absence of naked singularities on the would-be horizons.)

This then gives a small set of conditions that a spacetime must fulfil to have these nice features ---which is useful when analysing general stationary axisymmetric spacetimes, such as generalisations of the Kerr solution.

\bigskip
\hrule\hrule\hrule

\addtocontents{toc}{\bigskip\hrule}

\appendices
\section{Christoffel symbols}

Note the generic pattern of zero and non-zero entries in the Christoffel symbols:
\begin{equation}
\Gamma^r{}_{ab} \sim \Gamma^\chi{}_{ab} \sim  \left[\begin{array}{cc|cc}
*&*&0&0\\{}*&{}*&0&0\\ \hline 0&0&*&*\\0&0&*&*
\end{array}
\right];
\qquad\qquad
\Gamma^\phi{}_{ab} \sim \Gamma^t{}_{ab} \sim  \left[\begin{array}{cc|cc}
0&0&*&*\\0&0&*&*\\ \hline *&*&0&0\\{}*&{}*&0&0
\end{array}
\right].
\end{equation}
On horizon we have the more specific results:
\begin{equation}
\Gamma^r{}_{ab} \sim  \left[\begin{array}{cc|cc}
\infty&*&0&0\\{}*&{}0&0&0\\ \hline 0&0&0&0\\0&0&0&0
\end{array}
\right]
\quad
\Gamma^\chi{}_{ab} \sim  \left[\begin{array}{cc|cc}
\infty&*&0&0\\
{}*&*&0&0\\
\hline 0&0&*&*\\0&0&*&*
\end{array}
\right]
\qquad
\Gamma^\phi{}_{ab} \sim \Gamma^t{}_{ab} \sim  \left[\begin{array}{cc|cc}
0&0&\infty&\infty\\0&0&\infty&\infty\\
\hline{}\infty&\infty&0&0\\{}\infty&{}\infty&0&0
\end{array}
\right]
\end{equation}
Note the various infinities appearing here. This leads to potentially awkward $\infty\times 0$ contributions to the on-horizon geodesic equations. Ultimately this is why it is worthwhile to consider the on-horizon null geodesics as a limit of off-horizon null curves, allowing one to implicitly invoke the l'H\^opital rule when taking the on-horizon limit.

\section{Spherical symmetry}

In spherical symmetry one sets $\omega(r,\chi)\to 0$, and drops the $\chi$ dependence from 
$h_{rr}(r,\chi)$ and $h_{tt}(r,\chi)$. The $\chi$ dependence of $g_{\chi\chi}(r,\chi)$ and $g_{\phi\phi}(r,\chi)$ is very tightly constrained. In fact, without significant loss of generality (as long as one is not sitting at a wormhole throat) one can take
\begin{equation}
g_{tt}(r,\chi) \to e^{-2\Phi(r)} \Delta (r); \qquad g_{rr}(r,\chi) \to {1\over \Delta (r)};
\end{equation}
and
\begin{equation}
g_{\chi\chi}(r,\chi) \to {r^2\over1-\chi^2};\qquad g_{\phi\phi}(r,\chi) \to r^2 (1-\chi^2).
\end{equation}
Then
\begin{equation}
h_{tt}(r,\chi) \to e^{-2\Phi(r)}; \qquad h_{rr}(r,\chi) \to 1;
\end{equation}
and our general result reduces to
\begin{equation}
\kappa \to {1\over 2} \; e^{-\Phi(r_H)}\; \Delta'(r_H). 
\end{equation}
If we additionally choose to write $\Delta(r) = 1-2m(r)/r$ then
\begin{equation}
\kappa \to  e^{-\Phi(r_H)}\; {1-2m'(r_H)\over 2 r_H}. 
\end{equation}
This is as expected~\cite{3-function,DBH1}.

\section{Three-function distortion of Kerr}

In recent work~\cite{3-function} we had considered a 3-function extension of Kerr that exhibited some particularly desirable astrophysical properties (Hamilton--Jacobi separability of the geodesic equations, Klein--Gordon separability of the wave equation). 
We demonstrated that for these 3-function extensions of Kerr all horizons were automatically Killing horizons, and extracted a simple explicit form for the surface gravities. In comparing that formalism to the current framework we first note that for all the 3-function distortions of Kerr \cite{3-function} we have
\begin{equation}
N^2(r,\chi) \to {e^{-2\Phi(r)} [\Xi(r)^2 + a^2\chi^2] \Delta(r) \over
[\Xi(r)^2 + a^2]^2 + a^2 e^{-2\Phi(r)} \Delta(r) [1-\chi^2]}.
\end{equation}
Thence
\begin{equation}
h_{tt}(r,\chi) \to {e^{-2\Phi(r)}[\Xi(r)^2 + a^2\chi^2]\over [\Xi(r)+a^2]^2 + \O(\Delta)}; \qquad
h_{rr}(r,\chi) \to \Xi(r)^2 + a^2\chi^2;
\end{equation}
immediately leading to
\begin{equation}
\kappa \to {e^{-\Phi(r_H)} \; \Delta'(r_H) \over 2 [ \Xi(r_H)^2 + a^2 ]} .
\end{equation}
So the two analyses agree where they overlap.

\section*{Acknowledgements}

JB was supported by a Victoria University of Wellington PhD Doctoral Scholarship.
\\
MV was directly supported by the Marsden Fund, 
via a grant administered by the Royal Society of New Zealand.


\addtocontents{toc}{\bigskip\hrule}

\null
\vspace{-50pt}
\setcounter{secnumdepth}{0}
\section[\hspace{14pt}  References]{}
%

\end{document}